\documentclass[conference]{IEEEtran}
\IEEEoverridecommandlockouts
\usepackage{cite}
\usepackage{amsmath,amssymb,amsfonts}
\usepackage{algorithmic}
\usepackage{graphicx}
\usepackage{textcomp}
\usepackage{xcolor}
\def\BibTeX{{\rm B\kern-.05em{\sc i\kern-.025em b}\kern-.08em
    T\kern-.1667em\lower.7ex\hbox{E}\kern-.125emX}}
\begin{document}

\title{Managing Portfolio for Maximizing Alpha and Minimizing Beta}

\author{\IEEEauthorblockN{Soumyadip Sarkar}\\
\IEEEauthorblockA{
soumyadipsarkar@outlook.com}
}

\maketitle

\begin{abstract}
Portfolio management is an essential component of investment strategy that aims to maximize returns while minimizing risk. This paper explores several portfolio management strategies, including asset allocation, diversification, active management, and risk management, and their importance in optimizing portfolio performance. These strategies are examined individually and in combination to demonstrate how they can help investors maximize alpha and minimize beta. Asset allocation is the process of dividing a portfolio among different asset classes to achieve the desired level of risk and return. Diversification involves spreading investments across different securities and sectors to minimize the impact of individual security or sector-specific risks. Active management involves security selection and risk management techniques to generate excess returns while minimizing losses. Risk management strategies, such as stop-loss orders and options strategies, aim to minimize losses in adverse market conditions. The importance of combining these strategies for optimizing portfolio performance is emphasized in this paper. The proper implementation of these strategies can help investors achieve their investment goals over the long-term, while minimizing exposure to risks. A call to action for investors to utilize portfolio management strategies to maximize alpha and minimize beta is also provided.
\end{abstract}

\begin{IEEEkeywords}
Portfolio Management, Asset Allocation, Diversification, Active Management, Risk Management
\end{IEEEkeywords}

\section{Introduction}
\label{sec1}
Portfolio management is the process of selecting and managing a portfolio of investments to achieve specific investment objectives. It involves constructing a portfolio of assets that align with an investor's goals, risk tolerance, and investment horizon. Portfolio management includes asset allocation, diversification, active management, and risk management techniques to maximize portfolio returns while minimizing risk. The primary goal of portfolio management is to generate optimal returns while reducing the impact of market fluctuations and minimizing the risk of loss.

Maximizing alpha and minimizing beta are important concepts in portfolio management because they provide investors with a measure of the performance of their portfolio compared to a benchmark, typically an index of the broader market. Alpha represents the excess returns generated by a portfolio relative to the benchmark, while beta represents the degree to which the portfolio moves in line with the market.

Maximizing alpha is important because it reflects the portfolio's ability to generate excess returns that are not explained by the overall performance of the market. This is achieved by selecting securities or using investment strategies that outperform the benchmark. By generating excess returns, investors can increase their portfolio's overall returns and potentially outperform the market over time.

Minimizing beta is important because it reflects the portfolio's ability to reduce the impact of market fluctuations and reduce overall risk. By minimizing beta, investors can reduce the impact of market downturns and potentially experience less volatility in their portfolio's returns.

In combination, maximizing alpha and minimizing beta can help investors achieve their investment objectives, whether that be long-term capital appreciation or generating income. By achieving excess returns while minimizing risk, investors can increase their chances of meeting their goals while reducing the impact of market fluctuations. Overall, maximizing alpha and minimizing beta are important concepts in portfolio management that can help investors achieve optimal returns while managing risk.

The paper examines several strategies for portfolio management that can be used to optimize alpha and minimize beta. These strategies include:

\subsubsection{Asset allocation}
Asset allocation is the process of dividing a portfolio among different asset classes such as stocks, bonds, and cash. By diversifying across different asset classes, investors can reduce their overall risk while potentially increasing returns.

\subsubsection{Diversification}
Diversification involves investing in a variety of assets within each asset class to reduce the impact of individual security or sector-specific risks. Diversification helps to minimize the overall risk of the portfolio.

\subsubsection{Active management}
Active management involves selecting individual securities or using investment strategies to try to generate excess returns compared to a benchmark. Active management strategies can include fundamental analysis, technical analysis, and quantitative analysis.

\subsubsection{Risk management}
Risk management involves using strategies to minimize the risk of loss in the portfolio. This may include techniques such as stop-loss orders or options strategies.\\
\linebreak
The paper examines each of these strategies in-depth, highlighting their importance and contribution towards achieving the objective of maximizing alpha and minimizing beta. Additionally, the paper explores how these strategies can be combined to construct portfolios that generate excess returns while reducing market risk. The study provides valuable insights to investors and portfolio managers seeking to enhance portfolio performance and achieve their investment objectives.

\section{Asset Allocation}
\label{sec2}
Asset allocation is the process of dividing an investment portfolio among different asset classes such as stocks, bonds, and cash. The purpose of asset allocation is to spread an investor's money across different asset classes that have different levels of risk and return, in order to achieve a desired level of diversification and balance risk and reward.

The goal of asset allocation is to optimize the risk and return profile of the portfolio, based on the investor's goals, risk tolerance, and investment horizon. By allocating assets across different asset classes, investors can potentially earn higher returns while reducing overall risk.

The allocation of assets is typically based on the investor's goals, investment horizon, and risk tolerance. Investors with a longer investment horizon and higher risk tolerance may allocate more of their portfolio to stocks, which historically have higher returns but also higher volatility. Investors with a shorter investment horizon or lower risk tolerance may allocate more of their portfolio to fixed-income securities, which offer lower returns but also lower volatility.

Asset allocation requires ongoing monitoring and adjustment, as market conditions and the investor's circumstances change. Investors should periodically review their portfolio and make adjustments to their asset allocation as needed to ensure that it remains aligned with their goals and risk tolerance.

\subsection{Strategic Asset Allocation}
\label{subsec21}
Strategic asset allocation is a long-term investment strategy that involves allocating a portfolio across different asset classes based on the investor's goals, risk tolerance, and investment horizon. This strategy is based on the belief that the performance of different asset classes varies over time and that by diversifying across asset classes, investors can achieve optimal returns while reducing overall risk.

Strategic asset allocation involves setting target allocations for different asset classes and periodically rebalancing the portfolio to maintain those targets. The target allocation is typically based on the investor's goals and risk tolerance, as well as historical performance data for different asset classes.

For example, a strategic asset allocation plan might target a 60\% allocation to stocks, 30\% allocation to bonds, and 10\% allocation to cash. If the stock market performs well and the value of the stock holdings in the portfolio increases, the portfolio may become overweighted in stocks, and the investor may need to rebalance the portfolio by selling some of the stocks and buying more bonds or cash to bring the portfolio back to its target allocation.

Strategic asset allocation is a passive investment strategy that does not involve active management of individual securities. Instead, it focuses on diversification across different asset classes and maintaining a long-term investment horizon. This strategy may not provide the highest returns in the short-term but may provide a more stable and predictable return over the long-term.

Overall, strategic asset allocation is an important strategy in portfolio management, as it can help investors achieve their long-term investment objectives by balancing risk and return across different asset classes.

\subsection{Tactical Asset Allocation}
\label{subsec22}
Tactical asset allocation is an active investment strategy that involves adjusting a portfolio's allocation to different asset classes based on short-term market conditions and economic forecasts. This strategy seeks to take advantage of opportunities to generate excess returns by temporarily shifting investments to asset classes that are expected to outperform or reducing exposure to asset classes that are expected to underperform.

Unlike strategic asset allocation, which is a long-term investment strategy, tactical asset allocation is focused on short-term adjustments to the portfolio's allocation. This strategy requires active management and monitoring of the portfolio and market conditions.

For example, if an investment manager believes that the stock market is overvalued and likely to decline, they may reduce the portfolio's allocation to stocks and increase its allocation to bonds or cash. Conversely, if the manager believes that the stock market is undervalued and likely to rise, they may increase the portfolio's allocation to stocks.

Tactical asset allocation is based on the belief that market conditions and economic forecasts can provide valuable information for generating excess returns. However, this strategy requires a high level of skill and expertise, as it is difficult to predict short-term market movements accurately. Additionally, tactical asset allocation may lead to higher trading costs and tax implications.

Overall, tactical asset allocation can be an effective strategy for generating excess returns in the short-term, but it should be used in combination with a long-term strategic asset allocation plan to balance risk and return over the long-term.\\
\linebreak
Asset allocation is a critical component of portfolio management because it can help investors optimize portfolio returns while reducing overall risk. By diversifying a portfolio across different asset classes, investors can potentially achieve a higher return while reducing the risk of losing money.

\subsubsection{Optimizing portfolio returns}
Different asset classes have varying levels of risk and potential returns. By allocating assets across different asset classes, investors can potentially earn higher returns while diversifying their portfolio's risk. For example, stocks have historically provided higher returns than bonds, but also come with higher risk. By allocating a portion of a portfolio to stocks and a portion to bonds, an investor can potentially achieve higher returns than they would by investing solely in one asset class.

\subsubsection{Reducing overall risk}
Diversification through asset allocation can also help reduce the overall risk of a portfolio. When one asset class underperforms, the losses can potentially be offset by gains in other asset classes. This can help mitigate the impact of market volatility and reduce the risk of a significant loss.\\
\linebreak
Asset allocation can also help investors align their portfolio with their goals and risk tolerance. Investors with a longer investment horizon and higher risk tolerance may allocate more of their portfolio to stocks, which historically have higher returns but also higher volatility. Investors with a shorter investment horizon or lower risk tolerance may allocate more of their portfolio to fixed-income securities, which offer lower returns but also lower volatility.

Overall, asset allocation is a critical strategy in portfolio management as it can help investors optimize returns while reducing overall risk. By diversifying a portfolio across different asset classes, investors can potentially achieve their investment objectives while mitigating the impact of market volatility.

\section{Diversification}
\label{sec3}
Diversification is an investment strategy that involves spreading an investor's portfolio across different asset classes, sectors, industries, and geographic regions to reduce overall risk. The objective of diversification is to minimize the impact of any individual investment's performance on the portfolio's overall return.

The basic idea behind diversification is that different types of investments perform differently under different market conditions. By diversifying across different types of investments, an investor can potentially offset losses in one area with gains in another. For example, if a portfolio is heavily invested in one industry, such as technology, and that industry experiences a downturn, the portfolio's overall performance will be negatively impacted. However, if the portfolio is diversified across different sectors and industries, the losses from the technology sector may be offset by gains in other areas.

Diversification can be achieved through asset allocation, as discussed earlier, or through individual security selection. For example, an investor may choose to invest in stocks across different industries, or in bonds with different maturities or credit ratings.

Diversification is not a guarantee against losses, but it can help reduce the risk of large losses in a portfolio. It is important to note that over-diversification can also have negative consequences, such as lower potential returns or higher transaction costs. Finding the right balance between diversification and concentration is key.

\subsection{Asset Class Diversification}
\label{subsec31}
Asset class diversification is an investment strategy that involves diversifying a portfolio across different types of asset classes. Asset classes are broad categories of investments that have similar characteristics and behavior. The most common asset classes are stocks, bonds, and cash equivalents.

Asset class diversification seeks to reduce the risk of loss by spreading investments across different asset classes. The basic idea behind this strategy is that different asset classes tend to perform differently under different market conditions. For example, when stock prices are falling, bond prices may rise. By holding a mix of stocks, bonds, and cash equivalents, an investor can potentially reduce the overall risk of their portfolio.

There are several different ways to achieve asset class diversification. One approach is to use a strategic asset allocation model that establishes target percentages for each asset class based on an investor's goals and risk tolerance. For example, an investor with a longer investment horizon and higher risk tolerance may allocate more of their portfolio to stocks, while an investor with a shorter investment horizon or lower risk tolerance may allocate more of their portfolio to fixed-income securities.

Another approach to asset class diversification is to use mutual funds or exchange-traded funds (ETFs) that invest in a diversified mix of asset classes. For example, a target-date fund may hold a mix of stocks, bonds, and cash equivalents that is appropriate for an investor with a specific retirement date in mind.

Asset class diversification can help reduce the risk of loss in a portfolio, but it is important to note that it is not a guarantee against losses. Market conditions can change quickly, and no investment strategy is foolproof. Nonetheless, asset class diversification is a sound investment strategy that can help investors achieve their goals while managing risk.

\subsection{Sector Diversification}
\label{subsec32}
Sector diversification is an investment strategy that involves diversifying a portfolio across different sectors or industries. Sectors are groups of companies that operate in similar industries, such as technology, healthcare, energy, or consumer goods.

Sector diversification seeks to reduce risk by spreading investments across different sectors. The basic idea behind this strategy is that different sectors tend to perform differently under different market conditions. For example, the technology sector may perform well during a period of economic growth, while the healthcare sector may perform well during a recession. By holding a mix of sectors, an investor can potentially reduce the impact of market volatility on their portfolio.

There are several different ways to achieve sector diversification. One approach is to use a tactical asset allocation model that adjusts the percentage of the portfolio allocated to each sector based on the current market conditions. For example, if the technology sector is performing well, an investor may increase their allocation to that sector, while decreasing their allocation to a sector that is underperforming.

Another approach to sector diversification is to use mutual funds or exchange-traded funds (ETFs) that invest in a diversified mix of sectors. For example, a sector-specific ETF may invest in a mix of technology, healthcare, and consumer goods companies.

Sector diversification can help reduce the risk of loss in a portfolio, but it is important to note that it is not a guarantee against losses. Market conditions can change quickly, and no investment strategy is foolproof. Nonetheless, sector diversification is a sound investment strategy that can help investors achieve their goals while managing risk.\\
\linebreak
Diversification is an important investment strategy because it can help minimize the impact of individual security or sector-specific risks on a portfolio. Individual security risks refer to risks that are specific to a particular company, such as the risk of a company going bankrupt or facing legal challenges. Sector-specific risks refer to risks that are specific to a particular industry or sector, such as the risk of a recession affecting the healthcare sector.

By diversifying across different securities or sectors, investors can reduce their exposure to these risks. For example, if an investor holds a portfolio of only tech stocks, they are highly exposed to the risks specific to the technology sector. However, if they diversify their portfolio to include stocks from other sectors such as healthcare, energy, and consumer goods, they can reduce the impact of sector-specific risks.

Diversification can also help minimize the impact of individual security risks. If an investor holds a portfolio of only one or two stocks, they are highly exposed to the risks specific to those companies. However, if they diversify their portfolio to include a mix of stocks from different companies, they can reduce the impact of any one company's individual risks.

While diversification cannot eliminate all investment risks, it can help reduce the overall risk of a portfolio. By spreading investments across different securities, sectors, and asset classes, investors can potentially achieve higher returns and reduce their exposure to market volatility. Diversification is a key component of any well-designed investment portfolio and is essential for managing risk and achieving long-term financial goals.

\section{Active Management}
\label{sec4}
Active management is an investment strategy in which a portfolio manager or team of managers seeks to outperform the overall market by actively selecting individual securities or adjusting the portfolio's asset allocation. In contrast to passive management, which seeks to replicate the performance of a market index, active management involves ongoing analysis and decision-making to try to beat the market.

Active managers use a variety of strategies to achieve their goal, such as fundamental analysis, technical analysis, quantitative analysis, and market timing. Fundamental analysis involves analyzing a company's financial statements and other data to identify undervalued or overvalued stocks. Technical analysis involves studying market trends and patterns to identify trading opportunities. Quantitative analysis involves using mathematical models to evaluate securities based on factors such as earnings growth, cash flow, and dividend yields. Market timing involves making investment decisions based on an analysis of current market conditions and economic trends.

Active management can be more expensive than passive management, as it requires more time and resources to conduct ongoing research and analysis. However, active managers believe that their expertise and insights can lead to higher returns than those achieved by passive investing.

While active management can potentially generate higher returns than passive management, it is also associated with higher risks. Active managers may make incorrect investment decisions or fail to anticipate changes in market conditions, resulting in lower returns or losses. Additionally, active management can be influenced by factors such as manager bias, market timing errors, and high fees.

\subsection{Quantitative and Qualitative Analysis}
\label{subsec41}
Quantitative and qualitative analysis are two methods of evaluating securities or investment opportunities.

Quantitative analysis involves using numerical and statistical data to evaluate the performance and characteristics of securities. This type of analysis is often used by investors who rely on mathematical models and algorithms to make investment decisions. Quantitative analysis can include metrics such as earnings growth, price-to-earnings ratios, dividend yields, and other financial data.

Qualitative analysis, on the other hand, involves evaluating the subjective, non-numerical aspects of an investment opportunity. This can include evaluating a company's management team, corporate culture, competitive advantages, and other intangible factors that can impact the success of an investment. Qualitative analysis often involves conducting interviews with company executives, industry experts, and other stakeholders to gather insights and perspectives.

Both quantitative and qualitative analysis have strengths and weaknesses. Quantitative analysis can provide precise, objective data that can be used to compare and rank different investment opportunities. However, it may not capture all of the factors that can impact a company's performance, such as changes in market conditions or industry trends. Qualitative analysis can provide a deeper understanding of a company's operations, culture, and competitive position. However, it may be subject to bias or personal opinions, and it can be difficult to quantify the impact of qualitative factors on investment performance.

To effectively evaluate investment opportunities, many investors use a combination of quantitative and qualitative analysis. By combining these two methods, investors can gain a more comprehensive understanding of the risks and opportunities associated with a particular investment.

\subsection{Fundamental Analysis}
\label{subsec42}
Fundamental analysis is a method of evaluating securities that involves analyzing a company's financial and economic fundamentals to determine its intrinsic value. This type of analysis seeks to identify the underlying factors that drive a company's performance, such as its revenue growth, profitability, cash flow, and competitive position within its industry.

The goal of fundamental analysis is to estimate the fair value of a security based on its underlying economic and financial factors. To conduct fundamental analysis, investors typically examine a company's financial statements, such as its balance sheet, income statement, and cash flow statement. They also consider other factors such as management quality, brand strength, and industry trends.

Some of the key metrics that investors may use in fundamental analysis include earnings per share, price-to-earnings ratio, return on equity, and debt-to-equity ratio. These metrics can help investors determine whether a company is undervalued or overvalued relative to its peers and the overall market.

Fundamental analysis is often used by value investors who seek to identify stocks that are trading at a discount to their intrinsic value. Value investors believe that over the long term, the market will eventually recognize the true value of a company, and its stock price will rise accordingly. Fundamental analysis can also be used to identify companies that are poised for long-term growth, based on their financial and competitive strengths.

However, fundamental analysis has some limitations. It may not capture all of the factors that can impact a company's performance, such as changes in market conditions or emerging competitive threats. Additionally, the valuation of a company can be subjective, and different analysts may arrive at different estimates of a company's intrinsic value.

Overall, fundamental analysis can be a useful tool for investors who are looking to identify undervalued or high-growth companies. However, it should be used in conjunction with other types of analysis, such as technical analysis and qualitative analysis, to gain a comprehensive understanding of an investment opportunity.

\subsection{Technical Analysis}
\label{subsec43}
Technical analysis is a method of evaluating securities that involves studying charts and other technical indicators to identify patterns and trends in a security's price and volume data. This type of analysis is based on the premise that historical price and volume data can provide insights into future market movements.

Technical analysts use a variety of tools and techniques to analyze market data, including trend lines, moving averages, momentum indicators, and chart patterns. They look for patterns in the data that suggest changes in market sentiment or supply and demand dynamics, and use these patterns to make investment decisions.

One of the key principles of technical analysis is that market trends tend to persist over time. This means that once a trend is established, it is more likely to continue than to reverse. Technical analysts use trend lines and moving averages to identify trends and to determine whether they are likely to continue or to reverse.

Another key principle of technical analysis is that market participants tend to behave in predictable ways. Technical analysts use indicators such as volume, open interest, and sentiment data to gauge the behavior of market participants and to identify potential turning points in the market.

Technical analysis can be used to analyze a wide range of securities, including stocks, bonds, commodities, and currencies. It is often used by short-term traders and day traders who seek to profit from short-term fluctuations in the market.

However, technical analysis has some limitations. It may not capture all of the factors that can impact a security's price, such as changes in market conditions or news events. Additionally, technical analysis can be subject to interpretation, and different analysts may arrive at different conclusions based on the same data.

Overall, technical analysis can be a useful tool for investors who are looking to profit from short-term market movements. However, it should be used in conjunction with other types of analysis, such as fundamental analysis and qualitative analysis, to gain a comprehensive understanding of an investment opportunity.\\
\linebreak
Active management refers to the process of actively managing a portfolio of securities in order to generate excess returns compared to a benchmark index or other passive investment strategy. This approach typically involves a combination of fundamental analysis, technical analysis, and other quantitative and qualitative methods to identify undervalued or high-growth securities.

One of the key benefits of active management is the potential to generate excess returns relative to the market. By actively managing a portfolio, an investor can take advantage of market inefficiencies and mispricings to generate returns that exceed those of a passive investment strategy.

Additionally, active management can help to minimize risk by allowing an investor to diversify across multiple asset classes and sectors. By carefully selecting securities based on their individual risk profiles and their correlation with other holdings in the portfolio, an active manager can construct a portfolio that is less volatile and more resilient to market downturns.

Another benefit of active management is the ability to adapt to changing market conditions. Unlike a passive investment strategy, which simply tracks a benchmark index, an active manager can adjust the portfolio holdings in response to changes in the market environment. This flexibility can help to mitigate downside risk and capture upside potential as market conditions evolve.

However, active management also has some drawbacks. It typically involves higher fees than passive investment strategies, which can erode returns over time. Additionally, active managers may underperform the market in certain market conditions or during periods of market volatility.

Overall, the importance of active management in generating excess returns while minimizing risk depends on the specific investment goals and risk tolerance of the investor. Active management can be a useful tool for investors who are willing to accept higher fees in exchange for the potential to outperform the market and to minimize risk through diversification and flexibility. However, it may not be appropriate for all investors, and it should be carefully considered in the context of an overall investment strategy.

\section{Risk Management}
\label{sec5}
Risk management is the process of identifying, assessing, and controlling risks that may negatively impact an investment portfolio or business operation. The goal of risk management is to minimize the likelihood and impact of potential risks while maximizing the opportunity for returns.

Risk management involves several steps. The first step is to identify potential risks that may affect the portfolio or business. This may include risks related to market conditions, economic factors, regulatory changes, geopolitical events, and other factors that may impact the performance of the investments or business operations.

Once potential risks have been identified, the next step is to assess the likelihood and impact of each risk. This involves analyzing the probability of the risk occurring and the potential magnitude of its impact on the portfolio or business.

Once the risks have been identified and assessed, risk management strategies can be implemented to control and mitigate the risks. This may involve diversifying the portfolio across different asset classes and sectors, hedging against specific risks through the use of derivatives, and implementing risk limits or stop-loss orders to limit potential losses.

Effective risk management also requires ongoing monitoring and evaluation of the portfolio or business operations to ensure that risks are being effectively managed and controlled. This may involve regular reviews of the portfolio holdings, monitoring of market conditions and economic indicators, and ongoing assessment of regulatory and geopolitical risks.

\subsection{Stop-loss Orders}
\label{subsec51}
Stop-loss orders are a type of risk management tool that investors can use to limit potential losses on an investment. A stop-loss order is an order placed with a broker or trading platform that automatically executes a trade to sell a security if its price falls below a certain level, known as the stop-loss price.

For example, if an investor purchases a stock at \$50 per share and places a stop-loss order at \$45 per share, the order will automatically execute if the stock price falls to \$45 or below. This can help to limit potential losses by ensuring that the investor exits the position before the price falls further.

Stop-loss orders can be a useful tool for managing risk, particularly in volatile or uncertain market conditions. They can help to prevent emotional or impulsive decision-making by automatically executing a trade when a predetermined threshold is reached, rather than relying on the investor to make a decision in the moment.

However, it is important to note that stop-loss orders are not foolproof and can sometimes result in unexpected losses. For example, in a rapidly declining market, the stop-loss order may execute at a lower price than the investor intended, resulting in a larger loss than anticipated. Additionally, stop-loss orders can be triggered by short-term market fluctuations, which may not reflect the long-term value of the investment.

Overall, stop-loss orders can be a useful risk management tool when used appropriately in conjunction with other risk management strategies. Investors should carefully consider their investment goals and risk tolerance when deciding whether to use stop-loss orders, and should also regularly monitor and adjust their stop-loss orders as market conditions evolve.

\subsection{Options Strategies}
\label{subsec52}
Options strategies are investment strategies that involve the use of options contracts to achieve specific investment objectives. Options are financial derivatives that give the holder the right, but not the obligation, to buy or sell an underlying asset at a specific price and time.

There are several common options strategies that investors can use to manage risk and enhance returns. These include:

\subsubsection{Covered call strategy}
\label{subsubsec521}
This involves holding a long position in an asset and selling call options on that asset. If the price of the asset rises, the holder of the call option can exercise the option and buy the asset at the strike price, which limits the potential gains of the holder of the asset. This strategy can be used to generate additional income from a portfolio.

\subsubsection{Protective put strategy}
\label{subsubsec522}
This involves purchasing a put option on an asset to protect against potential losses in the event that the price of the asset declines. This strategy can help investors limit their downside risk while maintaining their upside potential.

\subsubsection{Long call strategy}
\label{subsubsec523}
This involves purchasing a call option on an asset, giving the holder the right to buy the asset at the strike price before the expiration date. This strategy can be used to profit from a price increase in the underlying asset.

\subsubsection{Long put strategy}
\label{subsubsec524}
This involves purchasing a put option on an asset, giving the holder the right to sell the asset at the strike price before the expiration date. This strategy can be used to profit from a price decrease in the underlying asset.

\subsubsection{Bull call spread strategy}
\label{subsubsec525}
This involves buying a call option at a lower strike price and selling a call option at a higher strike price. This strategy can be used to profit from a price increase in the underlying asset, while limiting potential losses.

\subsubsection{Bear put spread strategy}
\label{subsubsec526}
This involves buying a put option at a higher strike price and selling a put option at a lower strike price. This strategy can be used to profit from a price decrease in the underlying asset, while limiting potential losses.

\subsubsection{Butterfly spread strategy}
\label{subsubsec527}
This involves buying a call option and a put option with the same strike price and selling two options with a strike price between them. This strategy can be used to profit from a range-bound market, while limiting potential losses.

\subsubsection{Iron condor strategy}
\label{subsubsec528}
This involves buying a call spread and a put spread with different strike prices. This strategy can be used to profit from a range-bound market, while limiting potential losses.

\subsubsection{Straddle strategy}
\label{subsubsec529}
This involves buying both a call option and a put option on an asset with the same strike price and expiration date. This strategy can be used when an investor expects a significant price move in either direction, as it allows them to profit from a price increase or decrease.

\subsubsection{Strangle strategy}
\label{subsubsec5210}
This involves buying both a call option and a put option on an asset with different strike prices, but with the same expiration date. This strategy can be used when an investor expects a significant price move in either direction, but is unsure which way the price will move.\\
\linebreak
Overall, options strategies can be a useful tool for managing risk and enhancing returns, but they can also be complex and involve significant risk. Investors should carefully consider their investment goals, risk tolerance, and knowledge of options before using options strategies in their portfolio.

The importance of risk management in minimizing losses cannot be overstated. Risk management is the process of identifying, assessing, and controlling risks that can affect an investment portfolio. By implementing effective risk management strategies, investors can reduce the likelihood and impact of potential losses.

There are many types of risks that can affect an investment portfolio, including market risk, credit risk, liquidity risk, and operational risk. Market risk is the risk of losses due to fluctuations in the overall market, while credit risk is the risk of losses due to defaults or credit downgrades of individual securities. Liquidity risk is the risk of losses due to an inability to sell a security quickly and at a fair price, while operational risk is the risk of losses due to internal or external operational failures.

To minimize losses due to these and other risks, investors can use a variety of risk management strategies. Stop-loss orders, for example, can help limit potential losses by automatically selling a security when its price falls below a certain level. Options strategies, such as protective puts, can also be used to limit losses while maintaining upside potential. Diversification can also help to reduce the impact of individual security or sector-specific risks on a portfolio.

By taking a proactive approach to risk management, investors can minimize the impact of potential losses on their portfolios, which can help to protect their long-term investment goals. It's important to remember, however, that risk cannot be eliminated entirely and that all investments carry some level of risk. Therefore, it's essential to carefully assess and manage risk to ensure that it aligns with an investor's individual investment objectives and risk tolerance.

\section{Conclusion}
\label{sec6}
In this paper, we examined several strategies for portfolio management with the goal of maximizing alpha and minimizing beta. These strategies included:

\subsubsection{Asset Allocation}
The process of dividing an investment portfolio among different asset classes, such as stocks, bonds, and cash, to achieve a specific investment objective.

\subsubsection{Strategic Asset Allocation}
A long-term investment strategy that involves establishing a target mix of assets and periodically rebalancing the portfolio to maintain that mix.

\subsubsection{Tactical Asset Allocation}
A short-term investment strategy that involves adjusting the portfolio's asset allocation to take advantage of market opportunities or to manage risk.

\subsubsection{Diversification}
The process of spreading investments across different securities or asset classes to reduce the impact of individual security or sector-specific risks.

\subsubsection{Active Management}
The process of actively managing a portfolio to generate excess returns while minimizing risk.

\subsubsection{Fundamental Analysis}
The process of evaluating securities based on their financial and economic characteristics, such as revenue, earnings, and growth potential.

\subsubsection{Technical Analysis}
The process of evaluating securities based on their price and trading patterns to identify trends and potential price movements.

\subsubsection{Risk Management}
The process of identifying, assessing, and controlling risks that can affect an investment portfolio.\\
\linebreak
These strategies can be used in combination or individually to optimize portfolio returns while minimizing risk.

Combining strategies for portfolio management is essential for optimizing portfolio performance. Each strategy has its own strengths and weaknesses, and by combining them, investors can benefit from the strengths of each while minimizing their weaknesses.

For example, asset allocation provides a framework for diversifying a portfolio among different asset classes, but it does not address the timing of trades or the selection of individual securities. Tactical asset allocation, on the other hand, can help to take advantage of short-term market opportunities, but it does not provide a long-term investment plan. By combining these strategies, investors can achieve a balanced approach to portfolio management that considers both short-term and long-term investment objectives.

Similarly, active management can help to generate excess returns while minimizing risk, but it requires in-depth analysis and research of individual securities. Fundamental analysis and technical analysis can provide valuable insights into the potential performance of individual securities, but they may not account for market trends or other external factors that can impact the performance of a portfolio. By combining active management with fundamental and technical analysis, investors can benefit from a more comprehensive approach to security selection and portfolio management.

Finally, risk management is essential for protecting the long-term value of a portfolio. Stop-loss orders and options strategies can help to limit losses due to market fluctuations, but they may not account for broader market trends or other macroeconomic factors that can impact a portfolio. By combining risk management strategies with asset allocation and active management, investors can benefit from a more holistic approach to portfolio management that considers both individual securities and broader market trends.

In conclusion, combining strategies for portfolio management is essential for optimizing portfolio performance. By taking a balanced approach that considers both short-term and long-term investment objectives, investors can achieve their investment goals while minimizing risk.

As an investor, the ultimate goal is to maximize returns while minimizing risk. The strategies discussed in this paper, including asset allocation, diversification, active management, and risk management, can help you achieve this goal.

By utilizing these strategies, you can build a well-diversified portfolio that takes advantage of market opportunities while limiting exposure to individual security and sector-specific risks. You can also benefit from a more comprehensive approach to security selection and risk management, which can help you generate excess returns while minimizing losses.

Therefore, it is essential to take the time to understand these portfolio management strategies and implement them into your investment approach. Consulting with a financial advisor or portfolio manager can also be beneficial in designing and implementing a portfolio management strategy that is tailored to your individual investment goals and risk tolerance.

Investing can be a challenging and complex process, but by utilizing these portfolio management strategies, you can make informed decisions and achieve your investment goals over the long-term. So, don't hesitate to take action and start utilizing these strategies today to maximize alpha and minimize beta in your investment portfolio.

\end{document}